\documentclass{PoS}
\def\Om{\mbox{$\Omega_{\rm M}$}}
\def\OL{\mbox{$\Omega_{\Lambda}$}}

\title{Blazars and Gamma Ray Bursts}

\ShortTitle{Blazars and Gamma Ray Bursts}

\author{\speaker{Gabriele Ghisellini}\thanks{gabriele.ghisellini@brera.inaf.it}\\
        INAF -- Osservatorio Astronomico di Brera\\
        E-mail: \email{gabriele.ghisellini@brera.inaf.it}}

\abstract{
Blazars and Gamma Ray Bursts (GRBs) are the fastest objects
known so far.
The radiation we see from these sources originates in a jet 
of similar aperture angle, and we think it is the result
of the conversion of some of the jet kinetic energy into
random motion of the emitting particles.
Mechanisms for producing, collimating and accelerating the jets
in these sources are uncertain, and it is fruitful to
compare the characteristics of both class of sources in search
of enlightening similarities.
I discuss some general characteristics of blazars
and GRBs such as the power of their jets compared with what
they can extract through accretion, and the dissipation mechanism
operating in the jets of both classes of sources.
In both classes, there is a well defined trend between the 
bolometric power and the frequency at which this power 
is mainly emitted, but blazars are ``redder when brighter",
while GRBs are ``bluer when brighter".
Finally, I discuss some recent exciting prospects to use
blazars to put constraints on the cosmic IR--Optical--UV backgrounds,
and to use GRBs as standard candles to measure the Universe.
}

\FullConference{VI Microquasar Workshop: Microquasars and Beyond\\
		September 18--22 2006\\
		Societ\`a del Casino, Como, Italy}

\begin{document}

\section{Introduction}

It makes sense to compare blazars and Gamma Ray Bursts (GRBs),
looking for similarities and differences.
The bulk velocity in these systems differs from the light speed
by less than 0.5 per cent in blazars, and probably less than
0.005 per cent in GRBs.
In blazars, jet must be accelerated  without ``wasting" much of their 
power in radiation, while in GRBs the high initial internal
optical depth prevents us to
study in detail the starting phases of the jet acceleration process.
The central engine in both systems is not completely understood,
but the main ingredients should be the same: accretion onto a 
black hole that rapidly spins.

Consider also that GRBs are {\it not} explosive events. 
If their duration is directly linked to the activity of
the accretion, then they last at least for $10^6$ dynamical
times, and a factor 100 more if the "post--prompt" flares 
observed by Swift are produced by the central engine.
Their duration is then analogous to $\sim$3000 years
in the life of an AGN with a $10^8$ solar mass black hole.

Both classes of objects emit most of their radiation in
the $\gamma$--ray band: the radiation processes for blazars
and for the afterglow of GRBs are the same:
synchrotron and inverse Compton process, while some uncertainty
remains for origin of the prompt emission of GRBs.

Both classes of objects seems to form a sequence, as a function
of their power, which is linked to the overall spectral energy
distribution (SED), but the two sequences are opposite to each other:
blazar are bluer when dimmer, while bluer GRB are brighter.

Since the Universe is transparent to hard X rays and more opaque
to high--energy (GeV and TeV) $\gamma$ rays (since $\gamma$ rays
can be absorbed in $\gamma$--$\gamma \to$ e$^\pm$ collisions
with IR, optical and UV background photons), we can use the fact that
blazars and GRBs are strong X--ray and $\gamma$--ray emitters to
study the far Universe.
This is particularly true for blazars in the forthcoming GLAST era,
when it will be possible to study the optical--UV background
in a way similar to what done today for the IR background
using TeV emitting BL Lacs.
Finally, and more importantly,
the GRB sequence is so tight that it can be used to standardize the
GRB energetics and to use GRB as standard candles up to redshifts
unreachable by type Ia supernovae.

%
%

\section{Bulk Lorentz factors}

\subsection{$\Gamma=10$ in blazars?}

The most direct estimates of the bulk Lorentz factor $\Gamma$
in blazars come from VLBI observations of superluminal speeds
$\beta_{\rm app}$
(see e.g. Jorstadt et al. 2001 and Marscher, these proceedings).
One derives (since $\Gamma\ge \beta_{\rm app}$) values 
between 5 and 45 for FSRQ (powerful blazars with broad emission lines)
and much less for low power, TeV emitting BL Lacs
(see e.g. Edwards \& Piner 2002; Piner \& Edwards 2004; Giroletti et al., 2004).
In these very same sources, however, modeling of the SED
requires very large $\Gamma$--factors, especially when the TeV
flux is de--absorbed (see e.g. Krawczynski, Coppi \& Aharonian 2002; 
Konopelko et al. 2003;  Katarzynski et al. 2006).
Furthermore, the intra--day variability in the radio band,
if intrinsic, implies $\Gamma\sim 100$ (see Wagner \& Witzel 1995
for a review).
Even if due to interstellar scintillation, as it seems, 
the implied small angular size of the source still puts 
severe limits on its bulk Lorentz factor.
Theoretical limits on the maximum possible bulk Lorentz factor
($\Gamma<100$) have been derived by Begelman, Rees \& Sikora (1994),
while Ghisellini \& Celotti (2002) argued that the jet minimizes
its total power for a specific value of its bulk Lorentz factor.
For a specific source, the jet bulk Lorentz factor can vary
(see Marscher, this proceedings), bringing some support to the
``internal shock scenario" (see below).
There is some evidence and theoretical support to the idea that the
jet may be structured, having a fast spine
surrounded by a slower layer (with $\Gamma$ greater than $\sim$3
to avoid to see the counterjet; Laing 1993; Chiaberge et al. 2001;
Ghisellini, Tavecchio \& Chiaberge 2005).

Summarizing, there are some evidences that the typical values of
$\Gamma$ for blazars are larger then what thought previously, and
that different blobs in the same jet can have different $\Gamma$.
Also, the values of $\Gamma$ in low power jets is still an issue,
due to the largely different values derived from spectral modeling
(on scales of 0.01--0.1 parsecs) and VLBI data ($\sim$ 1 pc).

\subsection{$\Gamma=100$ in GRBs?}

The most common argument used to demonstrate the need of large 
bulk Lorentz factor in GRBs is the compactness argument.
This combines the most rapid variability timescales
(around 1--10 ms) with the observation, in a few bursts,
of GeV emission during the prompt phase (see e.g. the review by 
Fishman \& Meegan 1995).
If taken together, these two facts imply that the source,
if it is not moving relativistically, would be opaque
to the $\gamma$--$\gamma \to$ e$^\pm$ process, and no 
$\gamma$ rays would be seen.
The required $\Gamma$ to let the source be transparent
to the photon--photon process is $\Gamma>100$.
On the other hand the GeV emission may originate in the
afterglow, not during the prompt, and according to the
time analysis of Beloborodov, Stern \& Svensson (1998; 2000) 
the power spectrum has a break for timescales shorter than 0.1--1 
second, so the variability on 1--10 ms may not be indicative of 
the entire size of the emitting region. 
Therefore the limit $\Gamma>100$ becomes questionable, 
and one of the key issues that AGILE and GLAST are called
to clarify is the presence of high energy emission 
that can clearly be associated with the prompt phase of GRBs.
The other arguments to invoke a large $\Gamma$--factor
are theoretical and are somewhat model--dependent, being 
linked to the amount of baryon loading and the requirement
that the internal shocks thought to dissipate the carried
kinetic energy are more efficient increasing the contrast
between the $\Gamma$--factors of the colliding shells.

\section{Power}

The power of the jets in blazars is not known precisely,
although several ways to estimate it have been proposed.
The limiting factor are the still unknown matter
content of jet, the still unknown energy contained
in the baryon component of the radio lobes, and
also the fact that the radiation produced by the jet
is strongly boosted by beaming and then depends
on the not precisely known bulk Lorentz factor and
viewing angles.

For GRBs, the energetics of the received radiation is
accurately measured, but we have still a large uncertainty
concerning the efficiency of emitting this radiation.
The current estimates are $E_{\rm iso}=10^{50}$ -- $10^{54}$
ergs for the total emitted energy assuming isotropic emission.
The collimation corrected values are a factor 10--1000 less.

\subsection{Blazars}

Historically, the first estimate of the jet power 
was derived by estimating the energy content
of the radio lobes in FR II radiogalaxies,
assuming an equipartition magnetic field and
cold protons (therefore the minimum energy conditions).
If one also estimates the lifetimes of these structures,
then one has the average jet energy required
to energize the lobes.
In this way one derives $10^{43}$--$10^{44}$ erg s$^{-1}$ for FR I
radiogalaxies and $10^{46}$--$10^{47}$ erg s$^{-1}$ for FR II
radiogalaxies and radio--loud quasars.
Rawling \& Saunders (1991) noted that this power 
linearly correlates with the power in the narrow lines,
which have a factor 100 less power.
Since narrow line luminosities are also a factor $\sim 100$
less than the accretion disk luminosities, 
the jet power and the disk luminosities are of the same order.

One can also calculate the power carried by the jet by
inferring its density through modeling the observed SED
and requiring that the jet carries at least the particles
and the magnetic field necessary to make the radiation we see.
This has been done on the pc scale by Celotti \& Fabian (1993),
on sub--pc scale (the $\gamma$--ray emitting zone,
see e.g. Ghisellini \& Celotti 2002), 
and on the hundreds of kpc scale (the X--ray jets seen by Chandra, 
see e.g. Celotti, Ghisellini \& Chiaberge 2001; Tavecchio et al. 2000).
These studies suggest large values of the power transported
by the jet and require
the presence of a dynamically dominating proton component
(see also arguments by Sikora \& Madejski 2000).

The superb angular resolution of Chandra 
made possible to use two new methods:
the first is to calculate the energy needed to produce
the large scale jet X--ray emission (thought to be produced
by inverse Compton scattering with the microwave photons,
Tavecchio et al. 2000; Celotti, Ghisellini \& Chiaberge 2001), and
the second is to calculate the energy contained in the 
``X--ray bubbles" i.e. cavities though to be inflated by the jet
(see e.g. Allen et al. 2006).

The emerging picture is that the powerful jets (as the ones in FR II
radiogalaxies) remains relativistic
up to hundreds of kpc, and that therefore they dissipate only a 
small fraction of their power, which remains intact to power the radio lobes.
In these sources the jet power is of the same order of the accretion
luminosity, or even slightly larger occasionally (i.e. during flares).
Low power jets (as the ones in FR I radiogalaxies), even if they are born
with larger bulk Lorentz factor, can decelerate between the zone
of production of most of their radiation and the VLBI zone.
In these systems there are no or very weak broad emission lines, and
no sign of the thermal accretion disk luminosities, which is smaller
than the jet power.

A still hotly debated issue is what is the main energy carriers of jets.
Proton--electron, e$^\pm$ pairs or Poynting flux?
Part of the answer may come from observations in the X--ray band,
where a feature due to bulk Comptonization is expected 
(Begelman \& Sikora 1987;  Sikora, Begelman \& Rees 1994; 
Sikora et al. 1997; Moderski et al. 2004).
The spectrum of this feature has been calculated recently
(Celotti, Ghisellini \& Fabian, 2006) and has two components,
both with a quasi--blackbody shape: the first, produced by
the bulk Compton process between the cold jet material and
the photons coming directly from the accretion disk, is predicted to
be at UV energies and it is thus unobservable, while the second,
occurring in the X--ray band, is produced by comptonizing the
broad line photons and it can be observed if
other contributing processes, namely the self--Compton flux, 
do not hide it. 
The detection of this component would not immediately prove that the
jet is matter dominated (it could be made by initially cold e$^\pm$ 
pairs), but we could determine both the total number of leptons
and the $\Gamma$--factor. 
This information, together with other consideration
about the possible values of the magnetic field, would greatly
help in determining the real matter content.

\subsection{GRBs}

\begin{figure}
\vskip -1 true cm
\begin{tabular}{ll}
\hskip -1 true cm
\includegraphics[width=0.55\textwidth]{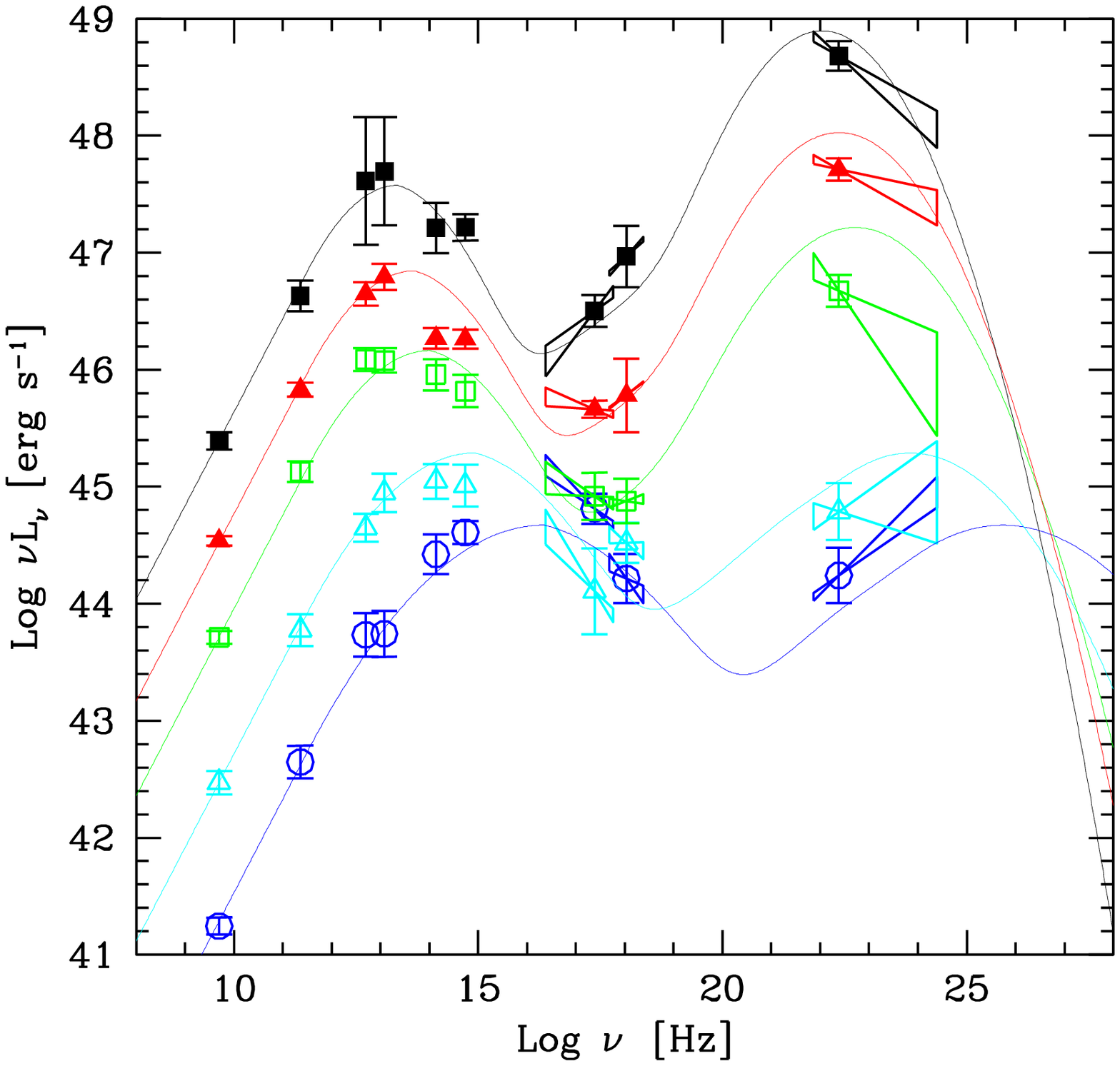}
%
&
\hskip -0.5 true cm
\includegraphics[width=0.55\textwidth]{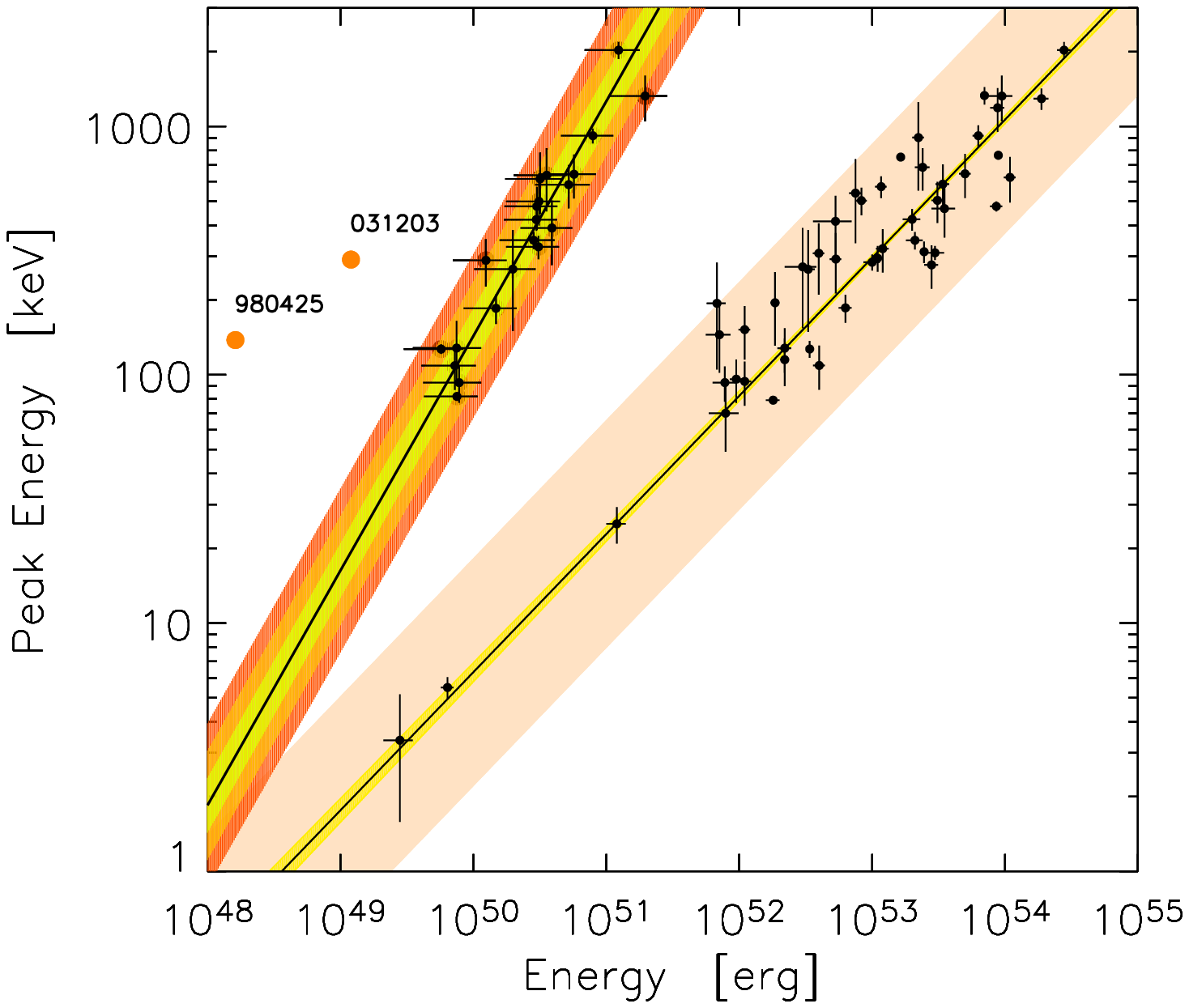}
\end{tabular} 
\caption{
{\bf Left:} The blazar sequence. 
The average SED of blazars belonging to different
bins of radio luminosity.
As can be seen, the radio luminosity tracks well
the bolometric one. 
More powerful blazars are redder, in the sense that the
peak frequencies of both the synchrotron and the
inverse Compton components have smaller values than in
low power blazars. From Fossati et al., (1998) and 
Donato et al. (2001).
{\bf Right:} The GRB sequences. 
The peak energy of the spectrum of the prompt emission 
(time integrated) of GRBs correlates both with the
isotropic bolometric energetics (on the right
of the figure, the so called ``Amati" relation",
$E_{\rm peak}\propto E_{\rm iso}^{1/2}$)
and with the collimation corrected energy (on the 
left, the so called ``Ghirlanda" relation, as
derived assuming that the circumburst density 
has a wind ($\propto r^{-2}$) profile.
In this case the correlation is linear.
Note that ``bluer" objects are brighter.
Adapted from Ghirlanda et al. 2004 and Nava et al. 2006.}
\label{fig1}
\end{figure}

Being transient events, for GRBs it is often more
meaningful to consider the total emitted energy instead
of their luminosities.
This can be calculated from the fluence (time integrated flux)
assuming isotropic emission, and the resulting $E_{\rm iso}$
values range between $10^{49}$ to $10^{54}$ erg (see Fig. \ref{fig1}).
The most energetic events emit approximately (if isotropic) 
the equivalent of one solar mass entirely transformed in energy.
Since there will be some efficiency $\eta$ involved, the total
(i.e. kinetic) energy of the jet must be larger by the factor $1/\eta$.
These huge values must be corrected if the emission is collimated.
There are indeed good evidences that this is the case,
because the lightcurve of the afterglow often shows a break
(at 0.2--5 days) which, if achromatic, is what is expected when
the bulk Lorentz factor $\Gamma$ (which is decreasing during the afterglow)
becomes equal to $1/\theta_{\rm j}$, the inverse of the jet opening angle.
Before this time, the observer sees only a fraction of the entire
jet surface (corresponding to an angle $1/\Gamma< \theta_{\rm j})$
while after this time the available emitting area is the entire jet surface.
The jet break time therefore measures $\theta_{\rm j}$ (through some
assumptions about the efficiency and the circumburst density).
The isotropic energetics can then be corrected by the solid
angle $(1-\cos\theta_{\rm j}$), and the resulting
energetics $E_\gamma$ (see Fig. \ref{fig1}) are distributed in 
a narrower range than  the isotropic ones, between $3\times 10^{49}$ 
and $10^{51}$ erg (Frail et al. 2001; Ghirlanda et al. 2004a).
Some independent constraints on the energetics
can be derived by the radio afterglow (see e.g. Frail, Waxman \& Kulkarni 2000
for GRB 970508), which can be thought as
the equivalent of the radio lobes in radiogalaxies, but unfortunately
these estimates could be made only for a few GRBs.

\section{Families}

\subsection{The blazar sequence}

Fossati et al. (1998), considering $\sim$100 blazars belonging to different
complete (flux limited) samples, constructed the average
SED for blazars in different bins of radio luminosities,
finding a well defined trend, illustrated in Fig. \ref{fig1}.
The SED is always characterized by two broad peaks,
which can be interpreted as due to the synchrotron and the 
inverse Compton processes.
The peak frequency of both peaks shifts to smaller values
(the blazar become redder) when increasing the bolometric luminosity.
At the same time, the high energy peak becomes more prominent.
Note also the ``valley" between the two broad peaks, located
in the X--ray band. 
This minimum of the SED strongly constrains the location
of the emission region, at least in powerful blazars with
broad emission lines, which reveals the presence of a
(normally) radiating accretion disk.
The argument is as follows: if the emitting region of the jet
is too close to the accretion disk, then the produced $\gamma$ rays
can interact with the X rays produced by the hot corona of the disk, 
producing e$^\pm$ pairs.
These pairs are born relativistic, and rapidly cool by inverse Compton
scattering the UV radiation of the disk. 
A significant fraction of this radiation would then fall in the ``valley",
contrary to what is observed.
As an order of magnitude, one can derive that the jet dissipation region
must be at least at $\sim100 R_{\rm s}$ (Schwarzschild radii) from the disk.
Note also that this argument apply only if the jet emission region
produces $\gamma$ rays: if for some reason they are not produced
(in a significant quantity), then the argument does not apply, and
we can have dissipation also close to the disk
(Katarzynski \& Ghisellini 2006).

The blazar sequence has been interpreted by Ghisellini et al. (1998)
as due to the different radiative cooling suffered by the
emitting electrons in sources of different powers.
In summary, powerful jets always live in systems where also
the broad emission lines are present and strong.
The line photons, used as seeds for the inverse Compton scattering,
efficiently cool the relativistic electrons, producing a break
(a cooling break) in the electron distribution at low energies.
Low power BL Lacs live in systems of no or weak emission lines:
only electrons of very high energy cool fast, and this produces
a break in the electron distribution at high energies.
The resulting SED is therefore bluer, with less power in the
high energy peak.

\subsection{The GRB sequences}

In very recent years, several correlations have been discovered,
linking the spectral properties of GRBs and their total energetics
or their peak luminosities (for a review see Ghirlanda, Ghisellini \& Firmani 2006).
All of them involve the energy at which the time integrated spectrum
peaks [in a $E F(E)$ diagram].
The fact that GRBs are ``bluer when brighter"
was indeed the surprising discovery after the finding 
(by Frail et al. 2001 among others), that the collimation corrected
energetics are more narrowly distributed than the isotropic ones.
In summary, the recently found correlations are
(the exponents are approximate):
\begin{itemize}
\item 
$E_{\rm iso}\propto E_{\rm p}^2$: the so called ``Amati" relation
(Amati et al. 2001; Amati 2006);
\item 
$L_{\rm p, iso}\propto E_{\rm p}^2$; the so called ``Yonetoku" relation,
(Yonetoku et al. 2004), where $L_{\rm p, iso}$ is the peak luminosity
of the prompt emission;
\item 
$E_{\gamma}\propto E_{\rm p}$: the so called ``Ghirlanda" relation,
where $E_\gamma =(1-\cos\theta_{\rm j})E_{\rm iso}$ is the collimation
corrected energetics.
It is linear if the circumburst density has a $r^{-2}$ 
(wind--like) profile (Nava et al. 2006), and it is of the form
$E_\gamma\propto E_{\rm p}^{3/2}$ if the circumburst medium is
homogeneous (Ghirlanda, Ghisellini \& Lazzati 2004);
\item 
$L_{\gamma}\propto E_{\rm p}^{1.2}$: where $L_\gamma$ is 
the collimation corrected peak luminosity
(Ghirlanda, Ghisellini \& Firmani 2006);
\item 
$E_{\rm iso}\propto E_{\rm p}^2 t_{\rm j}^{-1}$, where $t_{\rm j}$ is the time
when the afterglow light curve shows a steepening (Liang \& Zhang 2005);
\item
$L_{\rm p, iso}\propto E_{\rm p}^{3/2} T_{0.45}^{-1/2}$: 
the so--called ``Firmani" relation,
where $T_{0.45}$ is the time of enhanced burst emission, also used
for characterizing the variability behavior (Firmani et al. 2006).
\end{itemize}
In Fig. \ref{fig1} (right panel) we show the Amati relation and, 
on the same plot, the Ghirlanda relation in the ``wind case".
The derivation of the jet angle involves some assumptions, such as the
value of $\eta$ (assumed to be equal for all bursts) and the
density profile of the circumburst medium (homogeneous or wind--like).
Note that in the wind--like case, the $E_\gamma$--$E_{\rm p}$ correlation 
becomes linear (this is the correlation shown in Fig. \ref{fig1}; 
from Nava et al. 2006).
A linear correlation means that the number of photons in the peak
is the same for all bursts, and that the correlation is linear also in the
comoving frame of the bursts, since we transform both $E_\gamma$ and
$E_{\rm p}$ by the same factor $\Gamma$.
The equally tight correlation found recently by Firmani et al. (2006),
has the virtue of being model--independent and assumption free.
The tightest correlations (with reduced $\chi^2\sim 1$) are the Ghirlanda,
the Liang \& Zhang and the Firmani ones. 
The latter has also the virtue to relate quantities belonging to the 
prompt emission only, while the other two need to measure $t_{\rm j}$
by measuring the lightcurve of the afterglow.

There have been a few attempts to explain these correlations,
especially the Amati and the Ghirlanda ones.

Eichler \& Levinson (2005) considered them to be due to a viewing
angle effect, and consider a ``ring" geometry for the emission site.
In their view, all the bursts are intrinsically the same:
they have the same energetics and intrinsic $E_{\rm p}$.

Rees \& Meszaros (2005) pointed out that a tight correlation
between the energetics and the spectral peak suggest black body
emission, since in this case the number of free parameters is minimum.
They then suggest that the Amati relation can be obtained
if there is a correlation between the amount of dissipated power and the
location at which dissipation occurs.
Recently, Thompson (2006) and Thompson, Meszaros \& Rees (2006)
have suggested that the dissipation may instead occur at 
more or less the same location, thought as the radius of the 
progenitor star, and therefore of the order of $10^{10}$ cm
from the black hole.
The other ingredient is that, at this radius, the bulk Lorentz factor
should be of the order of $1/\theta_{\rm j}$.
This is justified by causality arguments, even if, in Thompson (2006)
the possibility to have a faster on--axis spine is mentioned.
This interpretation of the Amati correlation requires that most
of the observed energetics and the spectral peak corresponds
to a blackbody.
According to this idea, therefore, the observed spectra should
have a relatively strong blackbody component, especially visible
when analyzing time resolved spectra (the temperature may evolve in
time and the blackbody not be well visible in the time integrated spectrum).
Attempts to analyze the spectra this way have been done
(Ghirlanda et al. 20003; Ryde et al., 2005; Bosnjak et al. 2006)
and the decomposition of the time resolved spectrum as a sum
of a power law and a blackbody indeed yield acceptable results
when considering the BATSE energy range only (i.e. between
30 keV and $\sim$1 MeV).
However, the index of the fitted power law is in this case 
systematically softer than the low energy index ($\alpha$)
derived fitting a Band function, and the extrapolation to lower energies
becomes in conflict with the data of the Wide Field Camera
of $Beppo$SAX, for those bursts observed both by BATSE and $Beppo$SAX
(Ghirlanda 2006; Bosnjak et al. 2006 in prep.).

\section{Dissipation}

In 1978, Rees proposed that the jet of M87 could be powered
by collisions among different part of the jet itself,
moving at different speeds.
When colliding, they would produce shocks, giving rise to the
non--thermal radiation we see.
Although born in the AGN field, this idea of {\it internal shocks}
grew up more robustly in the GRB field, to become the
``paradigm" to explain their prompt emission
(see e.g. Rees \& M\'esz\'aros 1994).
Faster and later shells can then catch up slower earlier ones,
dissipating part of their bulk kinetic energy into radiation.
However, all shells are and remain relativistic: after the collision
the merged shells move with a bulk Lorentz factor which is
intermediate between the two initial ones.
It is then clear that this mechanism has a limited efficiency, because
only a small fraction of the bulk energy can be converted into radiation
(unless the contrast between the two initial Lorentz factor is huge,
see Beloborodov 2000; Takagi \& Kobayashi 2005).

This weak point for GRBs becomes instead a point in favor for
blazars: in fact, for them, we {\it require} a small efficiency,
for letting most of the bulk energy go un--dissipated to the
outer radio--lobes.
The other main point in favor for internal shocks in blazars is that 
it naturally explains why the main dissipation event, in the
relativistic jets, should occur at a few hundreds $R_{\rm s}$:
if two shells are initially separated by $R_0 \sim$ a few $R_{\rm s}$
and are moving --say-- with $\Gamma=10$ and $\Gamma=20$, they will
collide at $R_{coll}\sim R_0 \Gamma^2$, i.e. exactly where it is required.

A detailed study via numerical simulations of the predictions of
this model in blazars has been carried out by Spada et al. (2001)
for powerful objects (like 3C 279) and by Guetta et al. (2004) 
for less powerful objects, like Mkn 421 and BL Lac itself,
with encouraging results.
More recently, however, new ideas came out for the TeV BL Lac case,
involving jet deceleration (Georganopoulos \& Kazanas 2003) or
interactions between a fast jet spine and a slower layer (Ghisellini, 
Tavecchio \& Chiaberge 2005).

As discussed above, $R_{coll}\sim 10^2 R_{\rm s}$ when the typical 
$\Gamma$--factor of the shells is around 10.
What happens if the source produces, for some time, shells with
a smaller $\Gamma$?
This possibility has been explored by  
Katarzynski \& Ghisellini (2006), who assumed a jet working
at a constant average power, with shells of equal kinetic
energy $\Gamma Mc^2$ ($M$ is the mass of the shell).
Collisions of shells with low $\Gamma$ (and therefore large $M$)
occur closer to the jet apex than collisions between shells with
larger $\Gamma$.
This means that they dissipate when they are denser and more compact:
the magnetic field is larger, and the external radiation produced
by e.g. the broad line region, as seen
in the comoving frame, is less boosted.
All these facts means that the energy is radiated more through the
synchrotron process rather than by the Inverse Compton mechanism.
In turn, this means that fewer $\gamma$ rays are produced, and
even if they are partly transformed into pairs, they might not
``fill the valley" between the observed two broad peaks.
In this case we should observe a large (even dramatic) variability
in the IR--UV synchrotron and in the
X--ray self--Compton part of the spectrum, and much less 
in the MeV--GeV band produced by scattering broad line photons.
This is an ``economic" way to explain, for instance, the huge 
flare observed in the blazar 3C 454.3, in which the optical and the
X--ray flux increased by a factor 100 with respect to its ``normal"
level.
AGILE and GLAST will easily confirm or falsify these ideas.

\section{The maximum jet power}

If relativistic jets are powered (at least initially) 
by a Poynting flux, we can derive a simple expression for
the maximum value of their power, under some reasonable assumptions.
The Blandford \& Znajek (1977) power can be written as:
\begin{equation}
L_{\rm BZ}\, \sim \, 6\times 10^{20} 
\left({ a\over m}\right)^2 \left( {M_{\rm BH}\over M_\odot} \right)^2 B^2 
\,\,\,\, {\rm erg~s^{-1}}
\end{equation}
where $(a/m)$ is the specific black 
hole angular momentum ($\sim 1$ for maximally rotating black holes),
and the magnetic field $B$ is in Gauss.
Assume that the value of magnetic energy density
$U_B\equiv B^2/(8\pi)$ close to the hole is a fraction $\epsilon_B$ 
of the available gravitational energy:
\begin{equation}
U_B \, = \, \epsilon_B {GM_{\rm BH} \rho \over R} \,=\, 
\epsilon_B {R_s \over R} {\rho c^2 \over 2}
\end{equation}
The density $\rho$ is linked to the accretion rate $\dot M$ through
\begin{equation}
\dot M  \, = \, 2\pi RH \rho \beta_{\rm R}c
\end{equation}
where $\beta_{\rm R} c$ is the radial infalling velocity.
The mass accretion rate $\dot M$ is linked to the observed luminosity 
produced by the disk
\begin{equation}
L_{\rm disk} \, =\, \eta \dot M c^2
\end{equation}
The BZ jet power can then be written as:
\begin{equation}
L_{\rm BZ} \, \sim\, \left({ a\over m}\right)^2 { R_{\rm s}^3 \over R^2 H} 
{\epsilon_B  \over \eta} 
{L_{\rm disk} \over \beta_{\rm R} } 
\end{equation}
The maximum jet power is obtained setting $R\sim H\sim R_{\rm s}$ and
$a/m$, $\epsilon_B$, $\beta_{\rm R}$ equal to unity.
In this case
\begin{equation}
L_{\rm BZ, max} \, \sim\, {L_{\rm disk} \over \eta } \, =\, \dot M c^2
\end{equation}
This is in qualitative agreement with what can be estimated in blazars
and microquasars, and also in GRBs.
High power blazars (and therefore FR II radiogalaxies) can have
jets with a similar power than what emitted by the accretion disks, 
especially during (non--thermal) flares, when $P_{\rm jet}$ can even
be slightly larger.
For low power BL Lacs (and therefore FR I radiogalaxies) the luminosity
emitted by the accretion disk is largely sub-Eddington, and the efficiency
$\eta$ can be much smaller than the canonical 0.1 value.
This means that these systems can be largely jet dominated.
Of course the most extreme objects in this respect are GRBs,
in which most of the energy originated in the accretion process 
is advected into the hole, due to the huge densities and optical depths. 
According to the simple estimate given above, the most efficient
``engines" in the universe are jets, not accretion disks.

\section{Cosmological implications}

Hard X--ray photons are the ones suffering less from any absorption 
process in their travel from the sources to Earth, while higher 
energy photons can interact with IR--optical and UV photons and 
be absorbed through the $\gamma$--$\gamma \to$ e$^\pm$ process.
The fact that blazars and GRBs are strong emitters at both hard X--ray 
and $\gamma$--ray frequencies allow us to use them as torch-lights to 
study the amount of IR--UV cosmic backgrounds.
Furthermore, we have seen that the GRBs sequences (the Ghirlanda 
and the Firmani relations) can be used to standardize their energetics, 
enabling us to use them as standard candles.

\begin{figure}
\vskip -0.5 true cm
\begin{tabular}{ll}
\hskip -0.7 true cm
\includegraphics[width=0.55\textwidth]{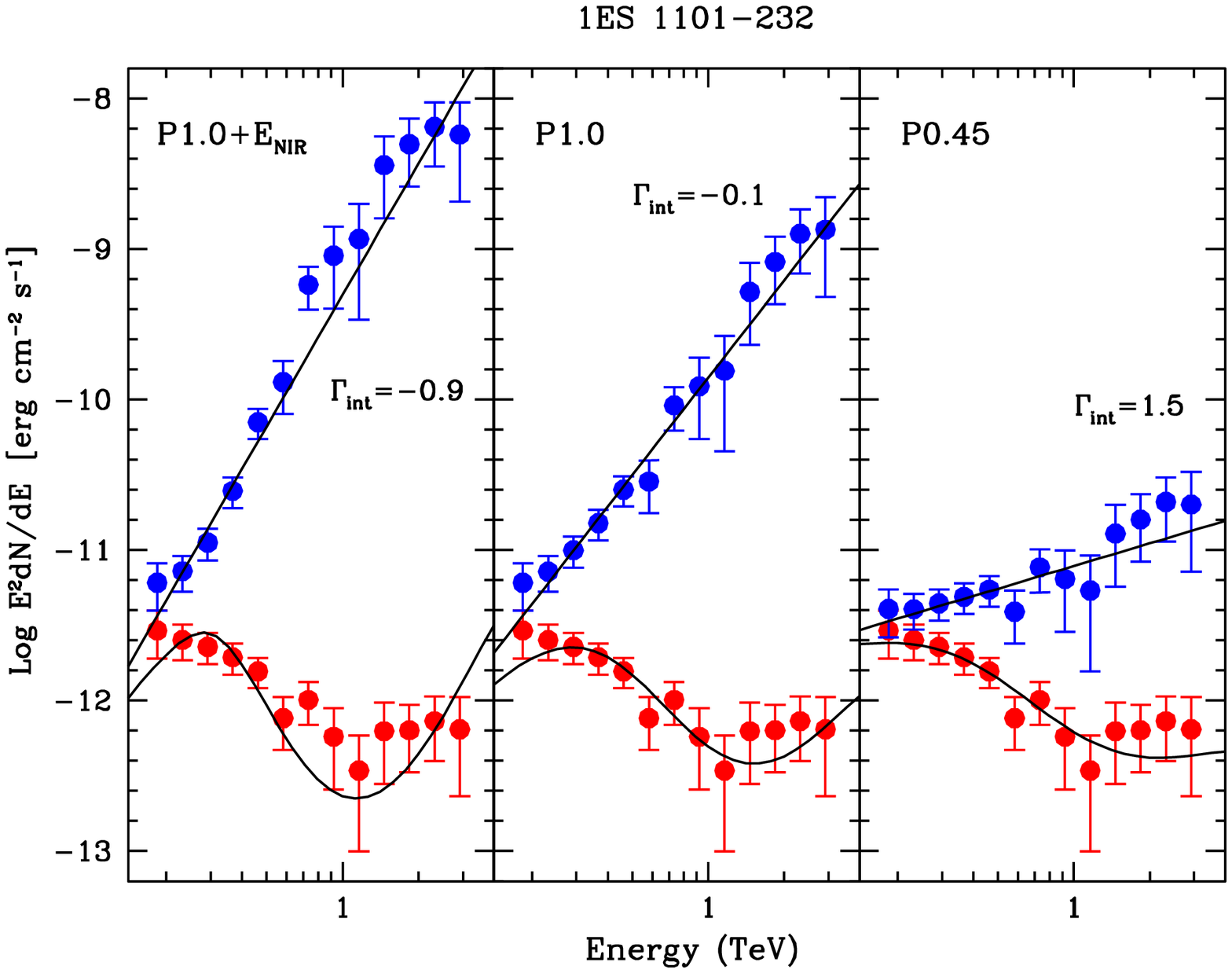}
&
\includegraphics[width=0.45\textwidth]{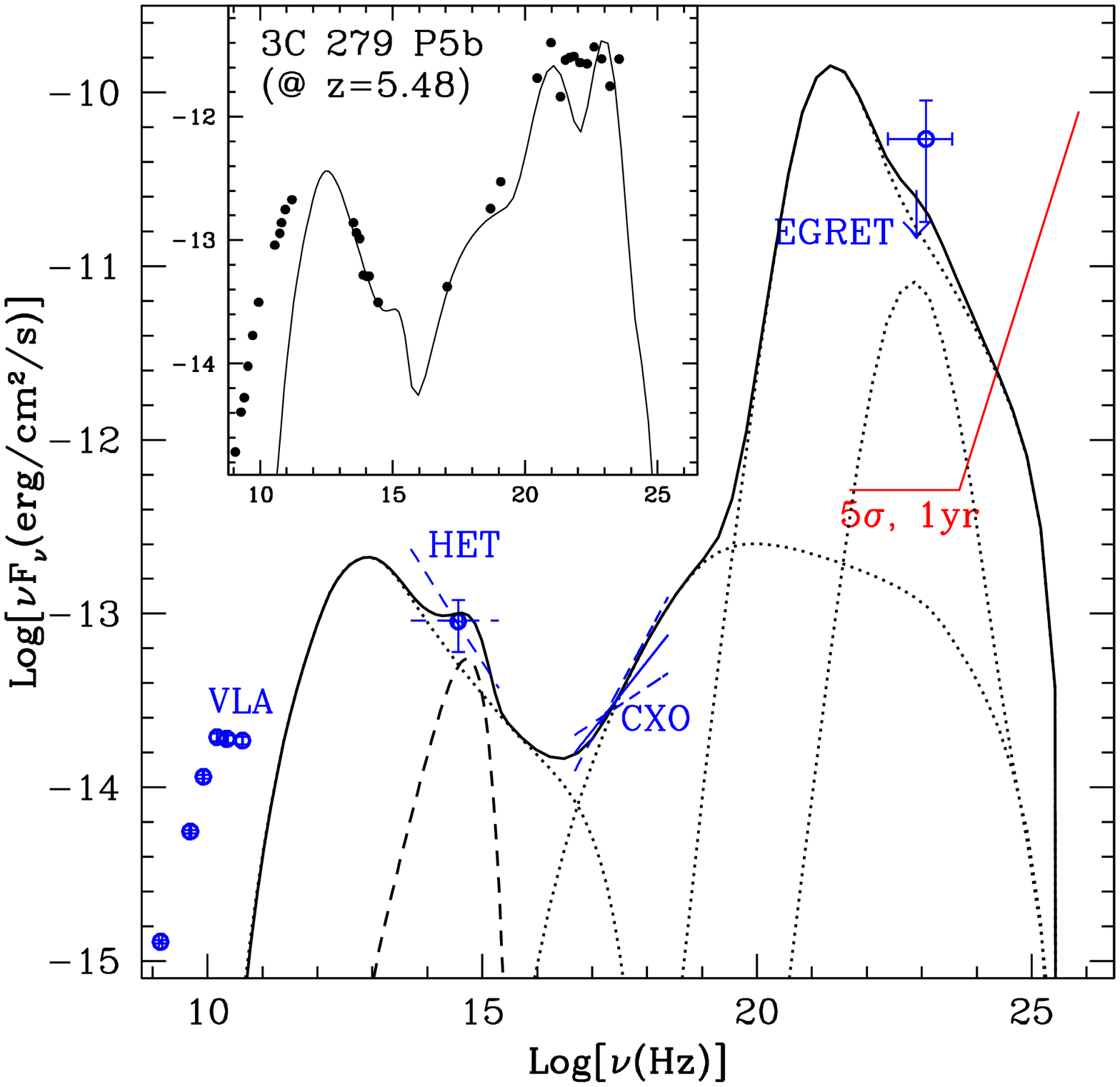}
\end{tabular} 
\caption{
{\bf Left:}
The TeV spectrum of 1ES 1101--232 (red points)
de--absorbed assuming different IR backgrounds
(blue points), decreasing from left to right.
The de--absorbed spectrum is very hard for the
left and central panel spectra.
The minimum possible IR background value 
corresponds to a an intrinsic photon spectrum
$\propto \nu^{-1.5}$ which is
considered by Aharonian et al. (2006) has
the most likely.
This conclusion was however questioned by
Katarzynski et al. (2006), who pointed out
that the limiting slope for the inverse Compton
photon spectrum, when the electron distribution has
a low energy cut--off, is $\propto \nu^{-2/3}$,
making stronger IR backgrounds still possible. 
(Figure from Aharonian et al. 2006.) 
{\bf Right:}
The SED of Q0906+6930. The solid line is a model.
Also shown is the limiting sensitivity of GLAST (red line),
to show that GLAST can well determine the $\gamma$--ray
spectrum  and any possible cut--off due to photon--photon
collisions. From Romani (2006).
}
\label{fig2}
\end{figure}

\subsection{Blazars}

Cherenkov telescopes have detected a dozen blazars in the TeV band, 
and EGRET detected $\sim$60 blazars in the GeV band.
Blazars, as a class, are as strong $\gamma$--ray emitters.
For the TeV emitting BL Lacs one of the key issues debated now is 
the determination of the poorly constrained cosmic IR background: 
we can use the absorption of their TeV photons as a tool to determine 
the level of the IR backgrounds. 
This however requires to assume an intrinsic spectrum,
which is model--dependent.
On the other hand, when we will have a sufficient number of sources 
distributed in redshift, it will be possible to determine the 
absorption cut--off energy in more model--independent way.
In Fig. \ref{fig2} we report the observed and the de--absorbed
TeV spectrum of the blazar 1ES 1101--232 (Aharonian et al. 2006).
Different IR backgrounds yield very different de--absorbed spectrum
and Aharonian et al. (2006) argued that to have a ``reasonable"
intrinsic spectrum (the one with the photon spectral index $\Gamma=1.5$)
it is necessary to assume the minimum possible IR background.
This conclusion was however questioned by
Katarzynski et al. (2006), who pointed out
that the hardest possible intrinsic spectrum
is the one having $\Gamma=-2/3$, corresponding to an
underlying electron distribution with a low energy cut--off.
In this case less extreme IR backgrounds are still possible.

In the AGILE and especially GLAST era the same 
study done now for the IR cosmic backgrounds can be done
also for the cosmic optical and UV backgrounds,
which are responsible for the absorption of GeV
photons (see e.g. Chen, Reyes \& Ritz 2004).
This means also to detect sources at relatively large
redshifts ($z>2$, because the Universe starts to 
be opaque to tens of GeV photons for these distances),
and to be able to detect a cut--off in their high
energy spectrum.
As a ``preview" of what GLAST can do for
cosmology using blazars, we show in Fig. \ref{fig2}
the SED of  Q0906+6930, the most distant blazars.
This source was recently associated by Romani et al. (2004)
to a previously unidentified EGRET source,
and it lies at the redshift $z=5.48$.
GLAST can likely find hundreds of blazars at redshifts
large enough to study the intervening absorption of
their GeV photons, once we can disentangle the absorption
made by the locally (i.e. accretion disk and broad line region)
produced optical--UV photons (Tavecchio et al. in preparation).

\subsection{GRBs}

The Ghirlanda, the Laing \& Zhang and the Firmani relations are 
tight enough to enable their use to standardize the energetics
of GRBs, and therefore we can use them as standard candles.
The obvious advantage is that GRBs are brighter than SN Ia,
therefore we can explore and measure the Universe 
beyond $z\sim 1.7$, the current redshift limit of SN Ia.
Furthermore, GRBs are free from extinction uncertainties
which may affect SN Ia.
There is however a problem: the correlations are found
using a particular cosmological problem, and we pretend
to use them to find the cosmological parameters.
This {\it circularity problem} has been dealt with in
several ways.

\begin{figure}
\vskip -1.5 true cm
\begin{tabular}{ll}
\hskip -1.5 true cm
\includegraphics[width=0.7\textwidth]{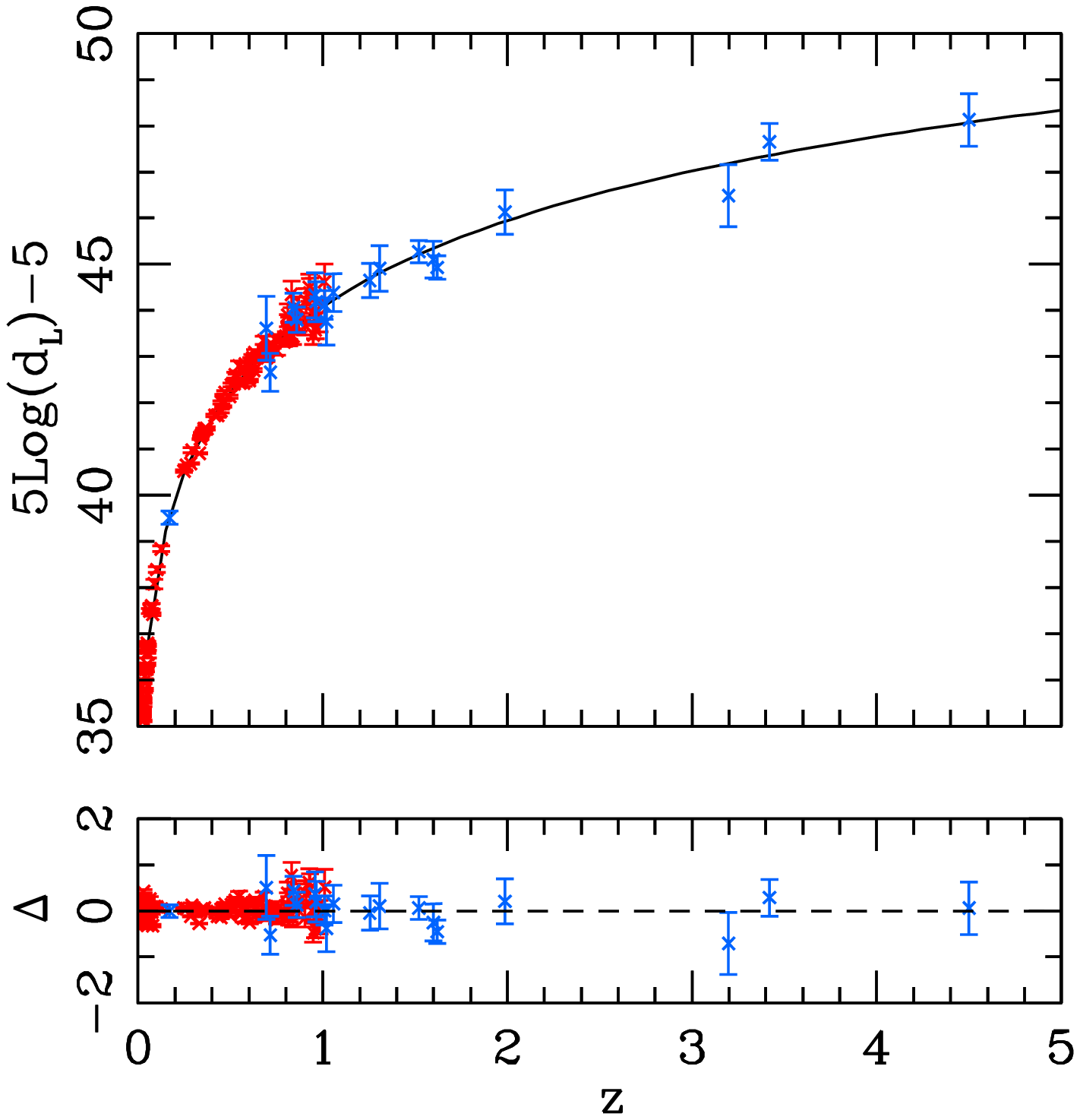}
%
&
\hskip -2.5 true cm \includegraphics[width=0.725\textwidth]{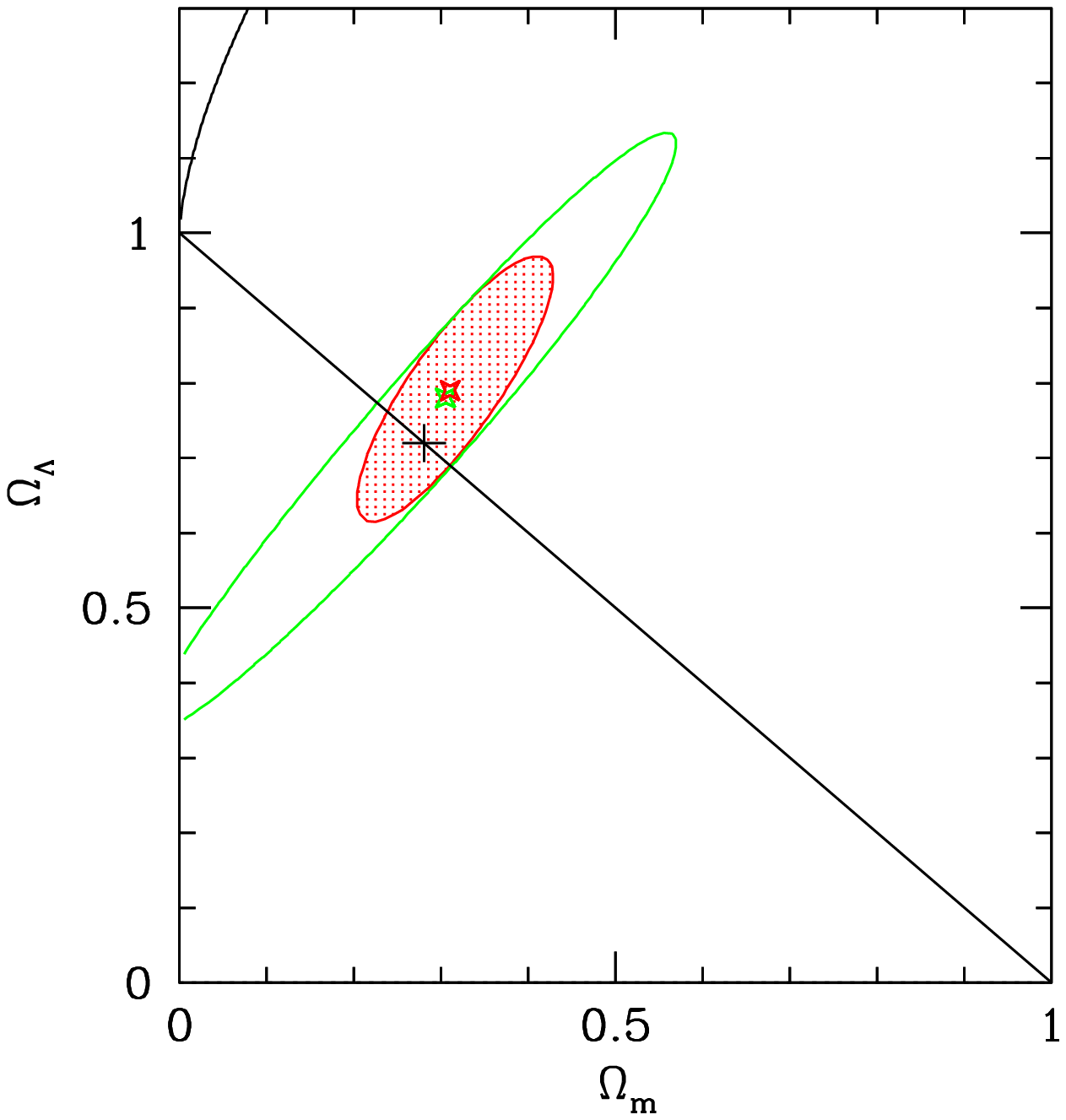}
\end{tabular}
\vskip -1 true cm 
\caption{
{\bf Left:} The Hubble diagram including SN Ia of the Legacy Survey
(red points)
and the GRBs analyzed in Firmani et al. (2006, blue points).
{\bf Right:}  The constraints on $\Omega_{\rm M}$--$\Omega_{\Lambda}$
set by considering SN Ia alone (green contours) and SN and GRBs
together (red shaded area).
It can be appreciated how much GRBs help in constraining
the cosmological parameters, and also the consistency with the
so--called "concordance cosmological model" (black cross).
(Both figures from Firmani et al. 2006b)
}
\label{fig3}
\end{figure}

\begin{enumerate}

\item
If we had a robust theoretical interpretation of the 
(Ghirlanda, Firmani or Laing \& Zhang)
correlation, then we could trust the theoretically derived slope and
normalization to use it for any choice of $\Omega_{\rm M}$
and $\Omega_{\Lambda}$
(see Dai et al. 2004 for what happens assuming
a fixed Ghirlanda correlation of the form $E_\gamma \propto E_{\rm p}^{3/2}$).

\item 
Ghirlanda et al. (2004b) proposed to use the scatter around the 
fitting line of the $E_{\rm peak}$--$E_\gamma$ correlation as
an indication of the best cosmology.
In other words, they find a given Ghirlanda correlation using
a pair of \Om, \OL, and measure the scatter of the points
around this correlation through a $\chi^2$ statistics.
Then they change \Om, \OL, find another correlation,
another scatter and another $\chi^2$ which can be compared 
with the previous one.
Iterating for a grid of \Om, \OL ~sets, they can assign to any point
in the \Om, \OL ~plane a value of $\chi^2$, and therefore draw 
the contours level of probability.

\item 
Firmani et al. (2005) proposed a more advanced method, 
based on a Bayesian approach, which is currently the best
to cure the circularity problem.
The basic idea is to take into account the information that 
the correlation is unique, even if we do not know its slope and normalization
(see Firmani et al. 2005 and Ghisellini et al. 2005 for the description 
of the method).

\item 
If one had many GRBs at low redshifts (i.e. $z<0.1$) the slopes
and normalizations of the correlations would be largely 
cosmology--independent, but there are not many nearby bursts.
On the other hand, the same goal can be achieved
using bursts in a narrow redshift bin: 
as suggested by Ghirlanda et al. (2006) and Liang \& Zhang (2006) 
a dozen of objects in a $\Delta z \sim 0.2$ around $z=1$ are sufficient.

\end{enumerate}

The other important point to be stressed is that GRBs should be considered
as {\it complementary}, not {\it alternative} to SN Ia.
Broadly speaking, one can think to the two classes of objects as standard candles
at large, located at low (SN Ia) and large (GRBs) redshifts.
The fact that also SN Ia need to be ``standardized" (through the  
``stretching" relation: more powerful SN Ia have lightcurves that decay
more slowly, see e.g. Phyllips et al. 1993) even requires a very similar
treatment for SN Ia and GRBs (even more so for the SN Ia of the 
Supernova Legacy Survey; Astier et al. 2006).

Applying the Bayesian approach to the Firmani relation, Firmani et al. (2006b)
already derived a very significant reduction in the uncertainties
in the \Om, \OL~ plane, when considering GRBs and SN Ia together.
This can be seen in Fig. \ref{fig3}: on the left there is the
classical Hubble diagram for SN Ia and GRBs, and on the right the
\Om, \OL~ constraints when considering GRB only (large elliptical
contours) and GRB+SN (smaller shaded area).
One can appreciate how the addition of GRBs can improve the
constraints, and also the fact that the joint fit is perfectly 
consistent with the so--called ``concordance cosmological model"
(i.e. \Om$\sim0.3$ and \OL$\sim 0.7$).

\section{Conclusions}

The values of the bulk Lorentz factor $\Gamma$ estimated for blazars,
have been increased during the years (except the $\Gamma$=100 peak reached 
when intraday variability was thought to be completely intrinsic).
Values of 40--50 are commonly proposed in the literature,
at least to explain $\gamma$--ray flares and especially for low power, 
TeV emitting, BL Lacs, and are also occasionally derived through 
the superluminal speeds of high frequency VLBI knots.
On the other hand the ``old" limits in GRBs (i.e. $\Gamma>100$)
coming from the compactness arguments rely on associating 
the high energy $\gamma$--ray emission to the prompt.
This is to be confirmed, because, instead, the high energy emission could 
belong to the afterglow.
For this issue AGILE and GLAST will be crucial.

Dissipation of the kinetic power of the jets into radiation
could happen at a variable distance from the black hole.
Internal shocks are natural candidates as a mechanism able
to transport bulk kinetic energy and to dissipate at the right place,
namely at 100--300 $R_{\rm s}$ in powerful blazars.

The sequences of blazars and GRBs are opposite: redder when brighter
for blazars, bluer when brighter for GRBs.
For blazars, the sequence has been explained by the increasing
dominance of the radiative cooling over the particle "heating"
as the luminosity increases.
For GRBs, it is less clear.

The power of these systems can exceed the luminosity of the 
accretion disk, and be of the order of $L_{\rm jet} \sim \dot M c^2$.

There are good prospects for the use of blazars and GRBs for cosmology.
For blazars, they are linked to the existing TeV Cherenkov observations
and the coming GeV--band AGILE and GLAST satellites.
Indeed, the use of distant blazars to measure the optical and UV
backgrounds can turn out to be one of the key programs for GLAST.
Also for GRBs GLAST can be crucial, due to its BATSE--like 
instruments, able to measure the peak energy in a reliable way
for a sufficient number of GRBs, whose location could be obtained
by a simultaneous detection by SWIFT or by ground based
robotic telescope images.

\vskip 0.3 true cm
{\bf Acknowledgements  ---} I thank Annalisa Celotti, Giancarlo Ghirlanda 
and Fabrizio Tavecchio for years of fruitful collaboration.

\end{document}